\documentclass[reprint,superscriptaddress,amsmath,amssymb,aps,prb,floatfix]{revtex4-2}
\usepackage{graphicx}
\usepackage{color}

\usepackage{dcolumn}
\graphicspath{{./Figures/}}
\usepackage{upgreek}
\usepackage{gensymb}
\usepackage{float}
\usepackage{bm}
\usepackage{xcolor}
\usepackage{physics}
\usepackage{longtable}
\usepackage{comment}
\usepackage{graphicx}
\usepackage{subfigure}

\usepackage{xr}
\makeatletter
\newcommand*{\addFileDependency}[1]{
	\typeout{(#1)}
	\@addtofilelist{#1}
	\IfFileExists{#1}{}{\typeout{No file #1.}}
}
\makeatother

\newcommand*{\myexternaldocument}[1]{
	\externaldocument{#1}
	\addFileDependency{#1.tex}
	\addFileDependency{#1.aux}
}

\myexternaldocument{ME_CFB_ver1_supps}

\begin{document}
	
\title{Magnon-phonon coupling of synthetic antiferromagnets in a surface acoustic wave cavity resonator}

\author{Hiroki Matsumoto}
\affiliation{Department of Physics, The University of Tokyo, Hongo, Tokyo 113-0033, Japan}

\author{Isamu Yasuda}
\affiliation{Department of Physics, The University of Tokyo, Hongo, Tokyo 113-0033, Japan}

\author{Motoki Asano}
\affiliation{NTT Basic Research Laboratories, NTT corporation, Atsugi, Kanagawa 243-0198, Japan}

\author{Takuya Kawada}
\affiliation{Department of Physics, The University of Tokyo, Hongo, Tokyo 113-0033, Japan}

\author{Masashi Kawaguchi}
\affiliation{Department of Physics, The University of Tokyo, Hongo, Tokyo 113-0033, Japan}

\author{Daiki Hatanaka}
\affiliation{NTT Basic Research Laboratories, NTT corporation, Atsugi, Kanagawa 243-0198, Japan}

\author{Masamitsu Hayashi}
\affiliation{Department of Physics, The University of Tokyo, Hongo, Tokyo 113-0033, Japan}
\affiliation{Trans-scale quantum science institute (TSQS), The University of Tokyo, Hongo, Tokyo 113-0033, Japan}

\newif\iffigure
\figurefalse
\figuretrue

\date{\today}

\begin{abstract}
We use a surface acoustic wave (SAW) cavity resonator to study the coupling of acoustic magnons in a synthetic antiferromagnet (SAF) and the phonons carried by SAWs.
The SAF is composed of a CoFeB/Ru/CoFeB trilayer and the scattering matrix of the SAW resonator is studied to assess the coupling.
We find that the spectral linewidth of the SAW resonator is modulated when the frequency of the excited magnons approaches the SAW resonance frequency.
Moreover, the linewidth modulation varies with the magnitude and orientation of the external magnetic field.
Such change in the spectral linewidth can be well reproduced using macrospin-like model calculations.
From the model analyses, we estimate the magnon-phonon coupling strength to be $\sim$15.6 MHz at a SAW resonance frequency of 1.8 GHz: the corresponding magnomechanical cooperativity is $\sim$0.66.  
As the spectral shape hardly changes in a CoFeB single layer reference sample under the same experimental condition, these results show that SAF provides an ideal platform to study magnon-phonon coupling in a SAW cavity resonator. 
\end{abstract} 

\maketitle

\begin{figure*}[hbt]
    \centering
    \includegraphics[width=1.0\linewidth]{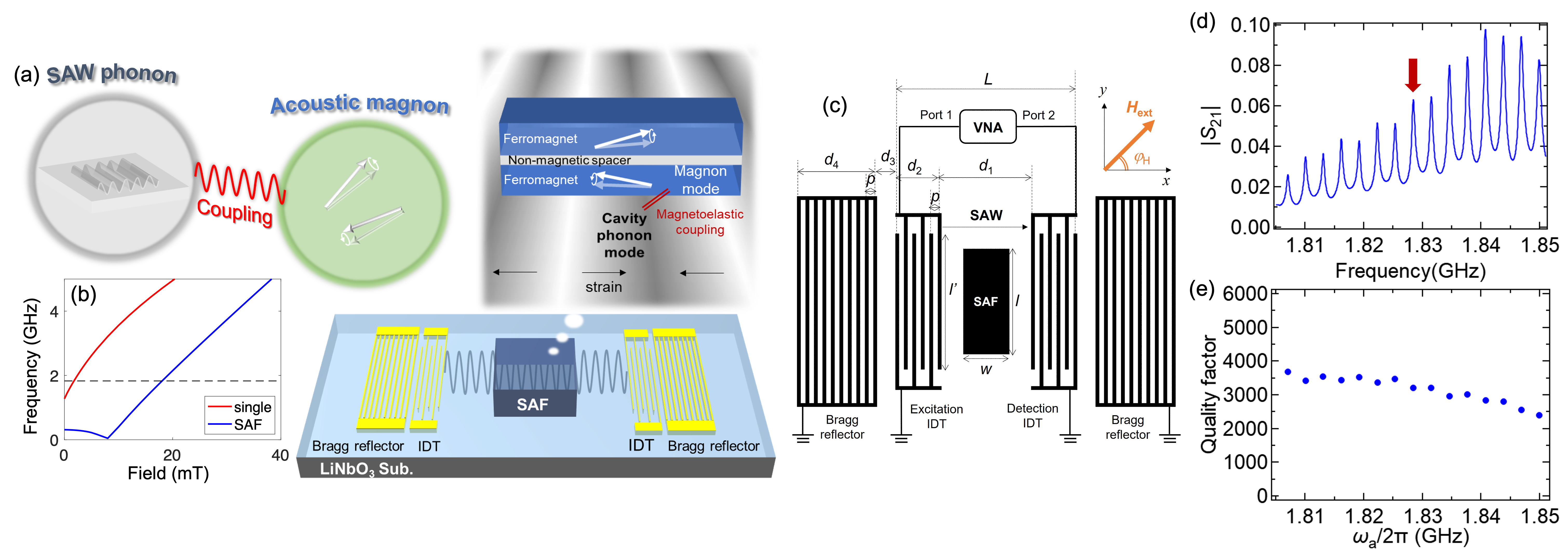}
	\caption{(a) Schematic illustration of the study. A synthetic antiferromagnet (SAF) is placed on the delay line of the SAW cavity resonator that is composed of the interdigital transducers and Bragg reflectors. SAW phonons in the cavity interact with acoustic magnons of the SAF. Phonon-magnon coupling is obtained by measuring the change in the SAW transmission spectra when the magnetic field magnitude and angle are varied. (b) Calculated magnon resonance frequency plotted as a function of in-plane magnetic field for a single layer film and a SAF. The horizontal dashed line represents the SAW resonance frequency used in the experiments. (c) Illustration of the device geometry. The design of the SAW cavity resonator is as follows: $L$ = 692 $\mu$m, $d_{1}$ = 618 $\mu$m, $d_{2}$ = 37 $\mu$m, $d_{3}$ = 2 $\mu$m, $d_{4}$ = 800 $\mu$m, $p$ = 1 $\mu$m, $w$ = 100 $\mu$m, $l$ = 600 $\mu$m, and $l$' = 695 $\mu$m. 
	The $x$-axis is parallel to SAW delay line and the film normal points along the $z$-axis. The relative angle between the $x$-axis and the external field is defined as $\varphi_\mathrm{H}$. (d) An exemplary transmission spectrum of the SAW cavity resonator. 
	The red arrow indicates the spectrum for which the results on the magnetic field dependence are presented in Figs.~\ref{fig2} and \ref{fig3}. 
	(d) Quality factor of the SAW transmission peaks as a function of the SAW resonance frequency $\omega_{a}/2\pi$. 
		\label{fig1}
	}
\end{figure*}

The interaction between phonons and magnons in magnetic systems is increasingly attracting interest not only for advancing understanding on the coupling mechanism\cite{weiler2011prl, dreher2012prb, thevenard2014prb, gowtham2015jap, an2020prb} but also for developing viable quantum technologies\cite{berk2019ncomm, li2021aplmater, hioki2022commphys}. 
Recent studies have shown that the cooperativity of the magnon-phonon coupled systems, a figure of merit that defines the degree of coupling, can be significantly larger than 1, achieving the necessary condition for the strong coupling\cite{xzhang2016sciadv, berk2019ncomm, hioki2022commphys}.
As the magnetostrictive interaction of the magnets is primarily responsible for the magnon-phonon coupling, systems with mechanical degrees of freedom are employed to facilitate the coupling.

Surface acoustic waves (SAWs) are acoustic phonons that propagate on the surface of piezoelectric substrate.
SAWs can be excited simply by applying a microwave electric field to an interdigital transducer (IDT) patterned on the substrate.
The simple device structure is attracting interest as it allows fabrication of on-chip phononic devices\cite{delsing2019jpd}.
Moreover, SAWs can propagate distances that exceed a millimeter\cite{casals2020prl}, exemplifying their exceptionally high coherence and small decay rate.
SAWs typically generate strains of the order of $10^{-6}$ and can reach $10^{-4}$ under appropriate conditions\cite{casals2020prl}.
Importantly, one can form a SAW cavity resonator by confining the waves in a region surrounded by Bragg reflectors, which are made of metallic electrodes\cite{bell1976ieee, manenti2016prb, bolgar2018prl, shao2019prap}.
The quality factor of such SAW cavity resonator can exceed $10^4$ at room temperature\cite{shao2019prap}.

Coherent excitation of magnons using SAWs in a ferromagnetic thin film has been reported since the pioneering work from Weiler \textit{et al}.\cite{weiler2011prl, dreher2012prb, li2021aplmater, sasaki2021ncomm, hatanaka2022prap}.
The resonance frequency to excite SAWs, defined by the period of the wires that comprise the IDT, is typically of the order of 1 GHz.
In contrast, the frequency of magnons in ferromagnetic thin films is in the range of a few GHz.
Previous studies have thus used materials with low magnetization, typically Ni, so that the resonance frequencies of the SAW and magnons match, a requirement for strong coupling.
However, even for Ni, the magnon resonance frequency is rather high such that the coupling can be studied only under a small magnetic field, which often causes multi-domain formation that impedes coherent excitation of magnons.
(Note that the quality factor of the SAW cavity resonator tends to decrease as the frequency increases\cite{safavinaeini2019optica}.)

To overcome this problem, one may use synthetic antiferromagnets (SAF) composed of two antiferromagnetically coupled ferromagnetic layers that are separated by a non-magnetic spacer layer\cite{parkin1990prl}.
The frequency to excite magnons in a SAF is typically smaller than that of its single layer counterpart (see Fig.~\ref{fig1}(b))\cite{kamimaki2020prap,shiota2020prl}, which in turn requires larger magnetic field to meet the condition of simultaneously exciting magnons and SAW.
Thus SAF will enable exploration of materials with large saturation magnetization, some of which possess small magnetic damping, and allow better control of the magnon-phonon coupling that would have been otherwise difficult in a single layer film.

Here we study the magnon-phonon coupling in a SAF composed of a CoFeB/Ru/CoFeB trilayer\cite{ishibashi2020sciadv,matsumoto2022apex,kuss2023prb} and a CoFeB single layer film, where the latter serves as a reference. 
We use a SAW cavity resonator to assess the coupling constant by extending the evaluation method previously developed\cite{hatanaka2022prap}.
We find that the spectral linewidth of the of the resonator is significantly modified only when acoustic magnons of SAF is excited; little change is found for the single layer film. 
The coupling constant, determined from the difference in the spectral linewidth with and without the magnon excitation, is $\sim$15.6 MHz.
As the change in the resonator spectra can be accounted for quantitatively using model calculations, these results show that SAF can be used as a platform to study the coupling of magnons and phonons for on-chip cavity magnomechanics.

Concept of this study and a schematic illustration of the device structure (SAW cavity with a trilayer SAF) are shown in Fig.~\ref{fig1}(a).
A pair of IDTs and Bragg reflectors composed of Ti(5)/Au(30) are fabricated on a piezoelectric Y+128$^\mathrm{o}$-cut LiNbO$_{3}$ substrate using a standard liftoff process that involves electron-beam lithography and vapor deposition.
The numbers in the parentheses show the layer thickness in nanometers.
Both widths and intervals of the wires in the IDTs and the Bragg reflectors are set to 500 nm.
A second liftoff process is employed to form a multilayer structure made of Ta(3)/Ru(3)/CoFeB(5)/Ru($t_\mathrm{Ru}$)/CoFeB(5)/Ru(3) with $t_\mathrm{Ru} = 0.53$ nm on the delay line (i.e. the path between the two IDTs) of the SAW cavity.
A single magnetic layer film made of 10 nm thick CoFeB ($t_\mathrm{Ru}$ = 0 nm in the film stacking) is made for comparison. 
RF magnetron sputtering is used to deposit the film and photolithography is used to pattern the device.
Details of the device structure are described in Fig.~\ref{fig1}(c).

The IDTs are connected to a vector network analyzer (VNA), which is used to carry out scattering matrix measurements of the device.
$S_{ij}$ with $i \neq j$ ($i = j$) represents the signal transmission (reflection) amplitude from port $j$ to port $i$ of the VNA ($i,j = 1,2$). 
An in-plane external magnetic field is applied during the scattering matrix measurements to change the magnetization direction of the CoFeB layer.
The relative angle between the magnetic field and the $x$-axis (parallel to the SAW propagation direction) is defined as $\varphi_\mathrm{H}$, as shown in the Fig. \ref{fig1} (c).
All measurements are performed at room temperature.

Figure~\ref{fig1}(d) shows an exemplary $|S_{21}|$ spectrum obtained from the scattering matrix measurements.
We find multiple peaks within the frequency range of 1.8 to 1.85 GHz.
Due to the SAW confinement within the cavity, the fundamental mode is split into the eigenmodes of the cavity.   
Using multi-peak fitting, we determine the acoustic resonance frequency $\omega_{a}/2\pi$, the acoustic damping $\kappa_{a}/2\pi$ (half width of half value) and the amplitude $A_a$ of each peak.
The following formula is used to fit the data:
\begin{equation}
\label{eq:fun:peak}
|S_{21}|^{2} = \sum_a \frac{A_a}{1 + (\frac{\omega - \omega_{a}}{2\kappa_{a}})^{2}} + \mathrm{const.},
\end{equation}
where $a$ runs over all the peaks observed in the $|S_{21}|$ spectrum.
We define the quality factor of each peak as $Q = \frac{\omega_{a}}{2 \kappa_{a}}$.
As shown in Fig.~\ref{fig1}(e), the quality factor of all peaks are larger than 10$^{3}$.
Such large quality factor ensures that the SAWs are effectively enclosed inside the cavity.
\begin{figure}[bt]
		\centering
		 \includegraphics[width=1.0\linewidth]{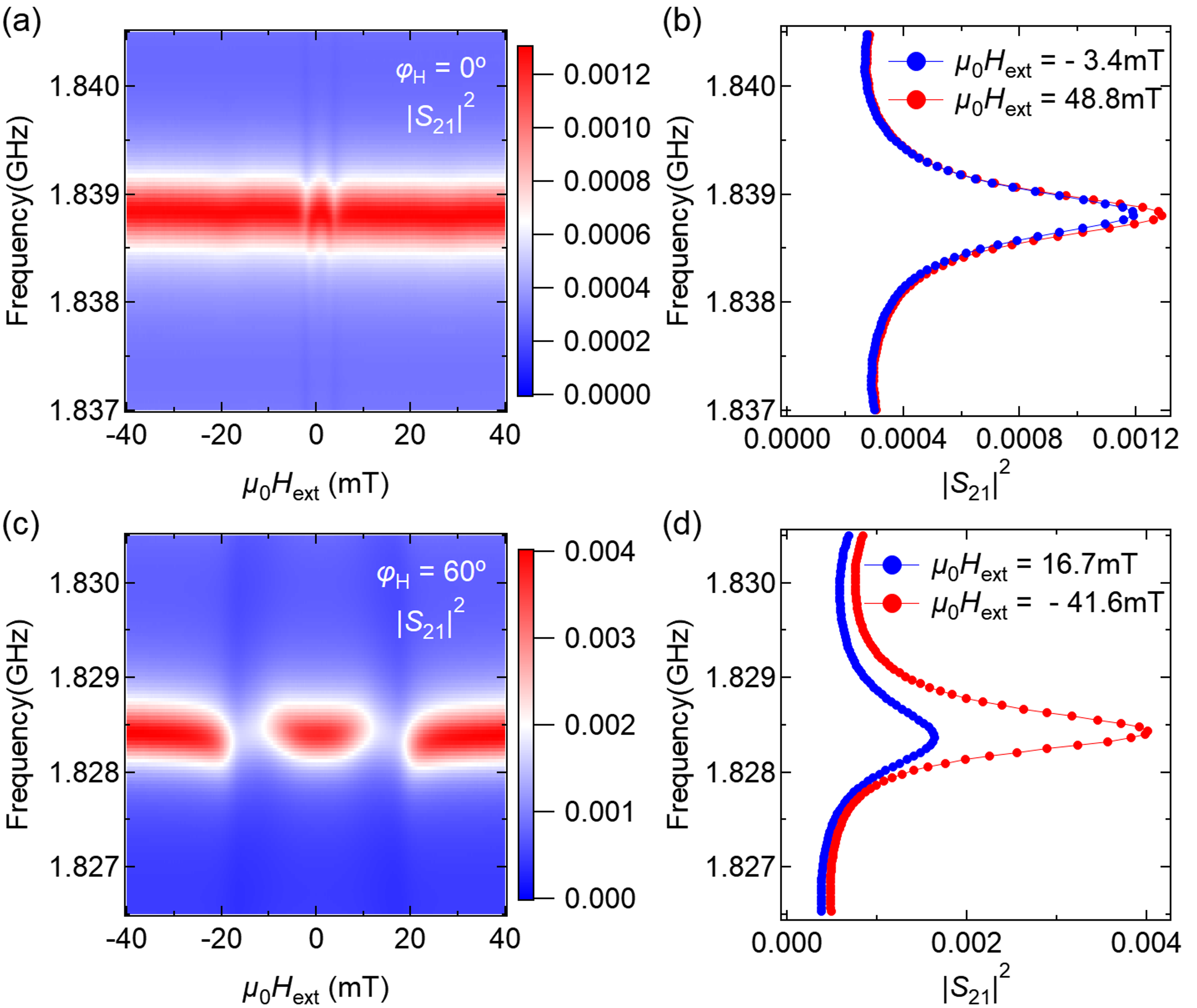}
	\caption{(a,c) Contour plot of $|S_{21}|^{2}$ as a function of the frequency and magnetic field measured for the CoFeB single layer film with $\varphi_\mathrm{H}$ = 0$^\mathrm{o}$ (a) and the SAF with $\varphi_\mathrm{H}$ = 60$^\mathrm{o}$ (c). 
	(b,d) The frequency dependence of $|S_{21}|^{2}$ obtained under constant $\mu_0 H_\mathrm{ext}$ for CoFeB single layer film (b) and SAF (d).
		\label{fig2}
	}
\end{figure}

We next show the change in the shape of the $|S_{21}|$ spectrum under application of external magnetic field for the single CoFeB layer film and the SAF.
Figure~\ref{fig2}(a) shows a contour plot of $|S_{21}|^{2}$ plotted as a function of the external magnetic field $H_\mathrm{ext}$ and the SAW excitation frequency for the single CoFeB layer film.
The magnetic field is applied along $\varphi_\mathrm{H}$ = 0$^\mathrm{o}$.
Here we show a contour plot of one of the multiple peaks found in the $|S_{21}|$ spectrum.
The color contrast represents the SAW transmission amplitude.
Near $\mu_0 H_\mathrm{ext} = 0$, small changes in the color contrast are found.
To show this more explicitly, line profiles of the spectrum along fixed $H_\mathrm{ext}$ ($\mu_{0}H_\mathrm{ext} = -3.4$ mT, 48.8 mT) are presented in Fig.~\ref{fig2}(b).
As evident, the peak amplitude is slightly reduced when $\mu_{0}H_\mathrm{ext} = -3.4$ mT compared to that of $\mu_{0}H_\mathrm{ext} = 48.8$ mT.

We have varied the magnetic field angle $\varphi_\mathrm{H}$ and obtained similar contour plots as shown in Fig.~\ref{fig2}(a).
For the CoFeB single layer film, we found notable change in $|S_{21}|^{2}$ only when $\varphi_\mathrm{H} = 0^\mathrm{o}$ and 180$^\mathrm{o}$, and the changes are small. 
In contrast, significant variation in $|S_{21}|^{2}$ with $H_\mathrm{ext}$ were found for the SAF film.
The largest change was observed when the field angle was set to $\varphi_\mathrm{H}=60^\mathrm{o}$: the results are presented in Fig.~\ref{fig2}(c).
Line profiles along fixed $H_\mathrm{ext}$ are shown in Fig.~\ref{fig2}(d).
Here we show plots along $\mu_{0}H_\mathrm{ext} =16.7$ mT and $-41.6$ mT.
Comparing the two plots, one finds not only the peak amplitude drops but also the peak frequency shifts and the linewidth broadens when $\mu_{0}H_\mathrm{ext} =16.7$ mT compared to those of $\mu_{0}H_\mathrm{ext} =-41.6$ mT.
Hereafter, we focus on the results of the SAF and discuss the results on the CoFeB single layer film later.

\begin{figure}[bt]
	\centering
	\includegraphics[width=1.0\linewidth]{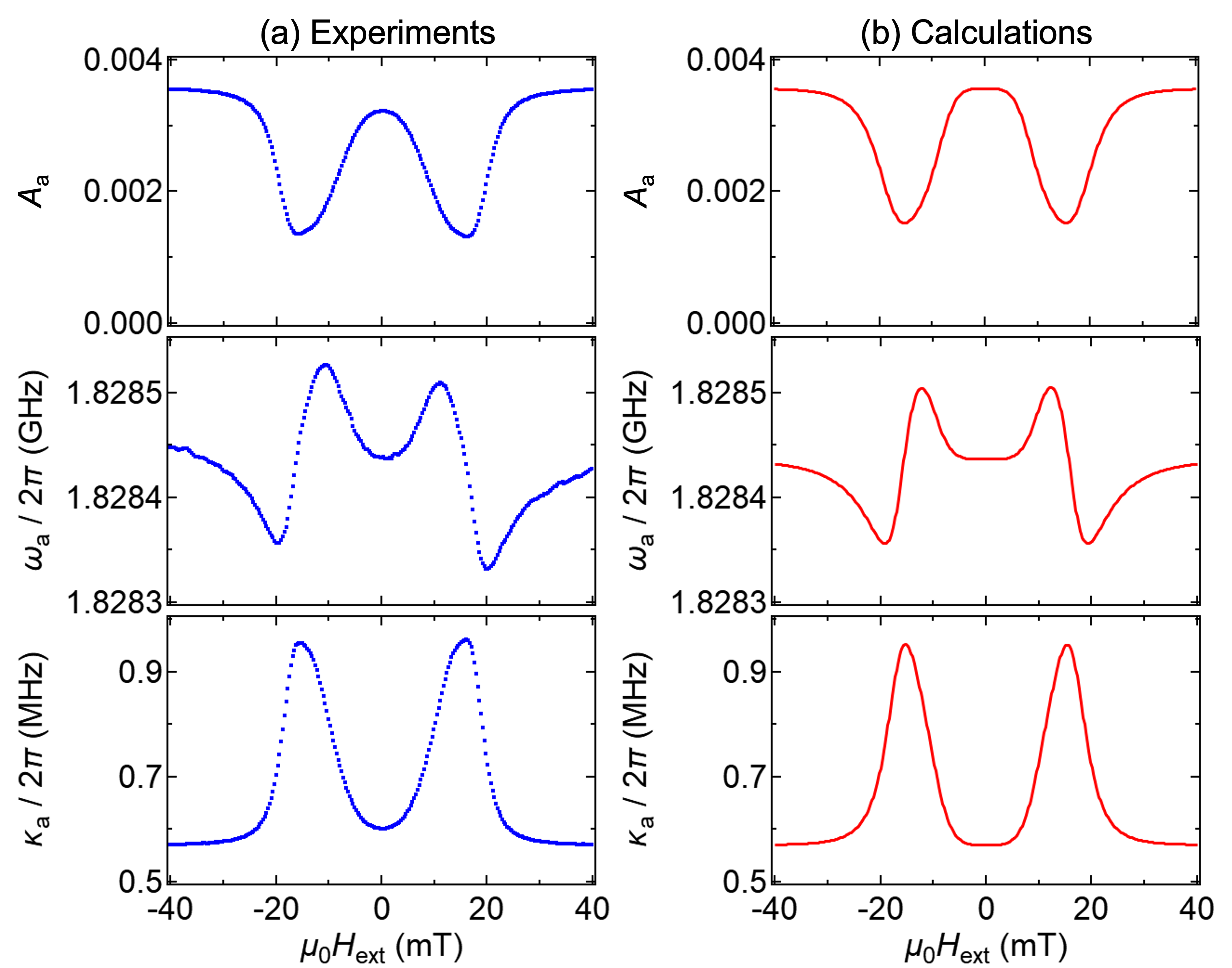}		
	\caption{The $H_\mathrm{ext}$ dependence of the $|S_{21}|^{2}$ transmission amplitude $A_a$ (upper), the peak frequency $\omega_{a}/2\pi$ (center) and the linewidth $\kappa_{a}/2\pi$ when the magnetic field is applied along $\varphi_\mathrm{H}$ = 60$^\mathrm{o}$. (a) experimental results, (b) calculations.
		\label{fig3}	}
\end{figure}

Figure~\ref{fig3}(a) shows the $H_\mathrm{ext}$ dependence of $A_a$, $\omega_{a}/2\pi$ and $\kappa_{a}/2\pi$ extracted from fitting the experimental results of the SAF (Fig. \ref{fig2}(c)) with Eq.~(\ref{eq:fun:peak}).
As evident, $A_a$ takes a minimum at $\mu_{0}H_\mathrm{ext} =16.7$ mT, suggesting that a fraction of the SAW power is absorbed.
$\omega_{a}/2\pi$ shows a significant change around $\mu_{0}H_\mathrm{ext} =16.7$ mT while $\kappa_{a}/2\pi$ takes a maximum.
Numerical calculations are performed to quantitatively account for these results.
The Landau-Lifshitz-Gilbert (LLG) equation and the elastic wave equation\cite{dreher2012prb, hatanaka2022prap, asano2023prb} are used to study the change in $A_a$, $\omega_{a}$ and $\kappa_{a}$ with $H_\mathrm{ext}$: see the Supplementary materials for the details.
Material parameters for the system are obtained from the experiments and the literature.
The calculation results are summarized in Fig.~\ref{fig3}(b).
Overall, we find the model calculations can account for the experimental results well.
From the calculations, we find that $\mu_{0}H_\mathrm{ext} =16.7$ mT is the resonance field, denoted as $H_\mathrm{res}$ hereafter, of the acoustic magnons in the SAF under an excitation frequency of 1.828 GHz and a field angle of $\varphi_\mathrm{H}=60^\mathrm{o}$.
(The resonance frequency of the optical magnons at the same field is $\sim$10.5 GHz.)
At this condition, the amplitude of the excited magnons becomes large such that they counteract on the SAW via the magnetostrictive coupling, causing the change in the SAW resonance frequency and linewidth around $H_\mathrm{res}$.
Such back action are often observed in coupled oscillator systems\cite{potts2021prx, hatanaka2022prap, asano2023prb}.
For large $|H_\mathrm{ext}|$, the acoustic magnons are not excited (off-resonance condition) and thus the SAW transmission properties hardly change. 

We estimate the strength of the magnon-phonon coupling using the change in $\kappa_{a}$ with $H_\mathrm{ext}$.
The coupled equations of motion of magnetization and the longitudinal stress associated with the SAW are linearized and Fourier transformed, from which we obtain the longitudinal stress amplitude $\mathcal{E}_{xx}$:
\begin{equation}
\begin{aligned}
\mathcal{E}_{xx} \approx \frac{\mathcal{F}_0}{- 2 \omega_a^0 \left( \omega - \omega_a^0 - \frac{\mathrm{Re}[ \chi_\mathrm{am} ]}{2 \omega_a^0}  \right) + i 2 \omega \left( \kappa_a^0 + \frac{\mathrm{Im}[ \chi_\mathrm{am} ]}{2 \omega} \right)},
\label{eq:coupled:2nd:fft:sol:3}
\end{aligned}
\end{equation}
where $\mathcal{F}_0$ is the excitation amplitude of the SAW, $\omega$ and $\omega_a^0$ are the SAW excitation and resonance angular frequency, $\kappa_a^0$ is the SAW decay rate. 
Note that $\omega_a^0$ and $\kappa_a^0$ represent quantities without the influence of magnon-phonon coupling.
$\chi_\mathrm{am}$ is defined as 
\begin{equation}
\begin{aligned}
\chi_\mathrm{am} \equiv \frac{i 2 \omega_a^0 g }{ i (\omega - \omega_m) + \kappa_m},
\label{eq:chiam}
\end{aligned}
\end{equation}
where $g$ is the the magnon-phonon coupling constant, $\omega_m$ and $\kappa_m$ are the resonance angular frequency and damping constant of the magnetization.
Equation~(\ref{eq:coupled:2nd:fft:sol:3}) shows that $\mathcal{E}_{xx}$ takes the form of a Lorentzian, consistent with the observed spectra shown in Fig.~\ref{fig2}(b,d) and Eq.~(\ref{eq:fun:peak}).
The peak position and the linewidth of the Lorentzian in Eq.~(\ref{eq:coupled:2nd:fft:sol:3}), defined as $\omega_a$ and $\kappa_a$, respectively, are modified under the influence of magnon-phonon coupling.
$\omega_a$ and $\kappa_a$ read
\begin{equation}
\begin{aligned}
\omega_a \equiv \omega_a^0 + \frac{\mathrm{Re}[ \chi_\mathrm{am} ]}{2 \omega} \approx \omega_a^0 + \frac{g (\omega_a^0 - \omega_m)}{(\omega_a^0 - \omega_m)^2 + \kappa_m^2},\\
\kappa_a \equiv \kappa_a^0 + \frac{\mathrm{Im}[ \chi_\mathrm{am} ] }{2 \omega} \approx \kappa_a^0 + \frac{g \kappa_m}{(\omega_a^0 - \omega_m)^2 + \kappa_m^2}.
\label{eq:omegaAkappaAdash}
\end{aligned}
\end{equation}
where we have substituted $\omega \approx \omega_a^0$ after the nearly equal symbol, the condition set in experiments.
We note that Eq.~(\ref{eq:omegaAkappaAdash}) is consistent with the results shown in Fig.~\ref{fig3}.
That is, $\omega_a$ and $\kappa_a$ are, respectively, antisymmetric and symmetric functions of $H_\mathrm{ext}$ with respect to $H_\mathrm{res}$.
This can be understood if one replaces $\omega_m$ and $\omega_a^0$ with $H_\mathrm{ext}$ and  $H_\mathrm{res}$
in Eq.~(\ref{eq:omegaAkappaAdash}), which is justified since $\omega_m$ scales with $H_\mathrm{ext}$ (see e.g. Fig.~\ref{fig1}(b)).

\begin{figure}[bt]
	\centering
	\includegraphics[width=1.0\linewidth]{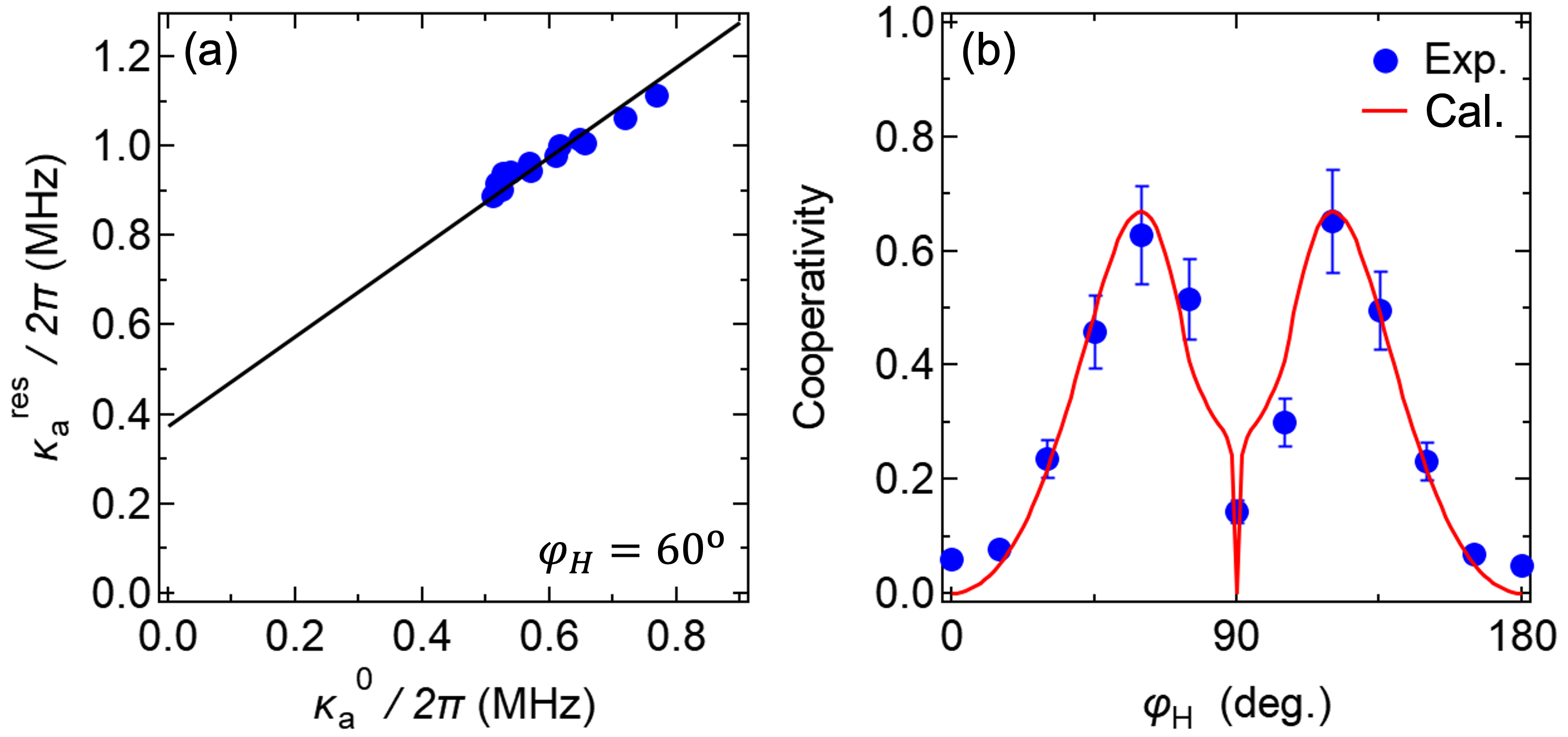}
	\caption{(a) The $\kappa_{a}^0$ dependence on $\kappa_{a}^\mathrm{res}$ in the SAF when $\varphi_\mathrm{H}$ = 60$^\mathrm{o}$. The blue circles and the black solid line show experimental results and linear fit to the data, respectively. The intercept of the linear line is equal to cooperativity parameter $C$. The slope of the linear line is fixed to 1. (b) The $\varphi_\mathrm{H}$ dependence of $C$ for the SAF. 
		\label{fig4}
	}
\end{figure}

To determine $g$, we study the difference in the linewidth $\kappa_a$ when the magnon frequency $\omega_m / (2 \pi)$, defined by $H_\mathrm{ext}$, is either set close to the SAW resonance frequency $\omega_a / (2 \pi)$ or set far from it. 
The linewidth of the former is defined as $\kappa_a^\mathrm{res}$ while that for the latter is equal to $\kappa_a^0$.
From the magnetic field dependence of $\kappa_{a}$ shown in Fig.~\ref{fig3}, we extract $\kappa_{a}^{\mathrm{res}}$ and $\kappa_{a}^0$.
The latter is obtained by taking the average $\kappa_{a}$ at large $H_\mathrm{ext}$ where the magnon-phonon coupling is suppressed.
We determine $\kappa_{a}^{\mathrm{res}}$ and $\kappa_{a}^0$ for all the peaks associated with the cavity eigenmodes shown in Fig.~\ref{fig1}(c).
Values of $\kappa_{a}^{\mathrm{res}}$ are plotted against $\kappa_{a}^0$ in Fig.~\ref{fig4}(a) with the blue circles.
As evident, $\kappa_{a}^{\mathrm{res}}$ linearly scales with $\kappa_{a}^0$.

To account for these results, we substitute the relation $|\omega_a^0 - \omega_m| \ll \kappa_m$, which applies to the system under investigation, into the second line of Eq.~(\ref{eq:omegaAkappaAdash}) to obtain
\begin{equation}
\begin{gathered}
\kappa_a^\mathrm{res} \approx  \kappa_a^0 \left( 1 + C \right),
\label{eq:kappa:C}
\end{gathered}
\end{equation}
where $C$ is the cooperativity defined as
\begin{equation}
\begin{gathered}
C \equiv \frac{g^2}{\kappa_a^0 \kappa_m}.
\label{eq:cooperativity}
\end{gathered}
\end{equation}
Equation~(\ref{eq:kappa:C}) shows the linear relation of $\kappa_{a}^{\mathrm{res}}$ with $\kappa_{a}^0$, consistent with the experiments.
We fit a linear function to the data presented in Fig.~\ref{fig4}(a) with the slope fixed to 1.
According to Eqs.~(\ref{eq:kappa:C}) and (\ref{eq:cooperativity}), the $y$-axis intercept of the linear fit gives $g^{2}/\kappa_{m}$, from which we find $g^{2}/(2\pi\kappa_{m})$ = 0.375 $\pm$ 0.006 MHz.
With $\kappa_a^0 / (2 \pi) = 0.570$ MHz (the average value obtained from all experiments), we obtain $C = 0.66$ from Eq.~(\ref{eq:cooperativity}).

The magnetic field orientation dependence of the cooperativity $C$ is shown in Fig.~\ref{fig4}(b).
$C$ takes a maximum at $\varphi_\mathrm{H}$ = 60$^\mathrm{o}$ and 120$^\mathrm{o}$.
Note that the magneto-elastic coupling is the largest when the relative angle between the magnetization and the strain (here it is parallel to the SAW propagation, i.e. along the $x$-axis) is $45^\mathrm{o}$ and $135^\mathrm{o}$.
The results shown in Fig.~\ref{fig4}(b) thus suggest that a non-zero magnetic anisotropy is present and influences the angular dependence of $C$.
To confirm this, the field angle dependence of the cooperativity is numerically calculated assuming an uniaxial magnetic anisotropy along the $y$-axis.
The results are shown by the red solid line in Fig.~\ref{fig4}(b).
The calculations are in good agreement with the experimental results.
The uniaxial magnetic anisotropy field of CoFeB is largely caused by the shape anisotropy: see Fig.~\ref{fig1}(c).

Finally, we determine the magnon-phonon coupling constant $g$ using the maximum $C$ found in Fig.~\ref{fig4}(a), which we denote as $C_\mathrm{max}$.
In order to estimate $g$ using Eq.~(\ref{eq:cooperativity}), one must find the value of $\kappa_{m}$.
We use the relation $\kappa_m \approx 2 \pi \alpha \gamma M_\mathrm{s}$\cite{asano2023prb} to obtain $\kappa_{m} / 2 \pi = 651$ MHz from the material parameters used in fitting the data presented in Figs.~\ref{fig3} and \ref{fig4}.
Substituting $\kappa_{m}$, $\kappa_a^0$ and $C_\mathrm{max} ( = 0.66)$ into Eq.~(\ref{eq:cooperativity}), we obtain $g/2\pi \sim$ 15.6 MHz.
This is similar to the value of $g$ reported previously for a Ni single layer film ($g/(2 \pi) \sim 9.9$ MHz), despite the difference in the product of the magneto-elastic coupling constant $b$ and saturation magnetization $M_\mathrm{s}$ that defines the magnitude of $g$.
Using reported material parameters, the calculated value of $M_\mathrm{s} b$ is larger for Ni than CoFeB, which cannot account for the values of $g$ found in the experiments.

Here we discuss the difference in the coupling constant $g$ for a single layer film and a SAF. 
In Fig.~\ref{fig1}(b), we show the calculated magnon resonance frequency as a function of an in-plane external magnetic field for a CoFeB single layer film and the CoFeB/Ru/CoFeB SAF.
Details of the calculations are presented in the supplementary material.
The horizontal dashed line shows the SAW resonance frequency.
For the single layer film, the magnetic field at which the magnon and the SAW resonance frequencies match is close to zero.
Under such condition, multidomain formation and magnetization switching of domains can occur and impede coherent magnon excitation.
This is corroborated by the fact that little change is found in the SAW resonance frequency and linewidth in the CoFeB single layer film (Fig.~\ref{fig2}(a,b)).
As we do not observe any signal that can be attributed to magnon excitation when the external field is oriented along $\varphi_H \sim 45 \pm 90 n$ deg ($n$: integer) where the magneto-elastic coupling is the strongest, we infer that the SAW cannot efficiently excite magnons under the experimental condition for the CoFeB single layer film. 

In contrast, the resonance frequency of acoustic magnons of SAF is reduced compared to the single layer film: see Fig.~\ref{fig1}(b). As a consequence, the magnetic field at which the acoustic magnon and SAW resonance frequencies coincide becomes sufficiently large such that multidomain formation and domain switching are suppressed. 
The fact that the magnetic field dependence of the spectral parameters (Figs.~\ref{fig3}) and the field angle dependence of the cooperativity (Figs.~\ref{fig4}(b)) can be well reproduced using model calculations\cite{asano2023prb} demonstrate that SAF provides an ideal platform to study magnon-phonon coupling.

In summary, we have studied the magnon-phonon coupling in a CoFeB/Ru/CoFeB tri-layer SAF and a CoFeB single layer film using a surface acoustic wave cavity resonator.
The cavity resonator exhibits a quality factor that exceeds $\sim 10^{3}$, providing strong confinement of the SAWs.
The SAW transmission spectra are measured under application of an in-plane magnetic field.
For the SAF, the SAW spectrum of the cavity eigenmodes exhibits reduction in amplitude as well as frequency shift and linewidth broadening at the resonance field of the acoustic magnons.
Interestingly, such changes in the SAW spectra are hardly found for a CoFeB single layer film.
We infer the difference is due to the difference in the magnon resonance field.
The resonance field is close to zero for the single layer film, where multidomain structures can form to impede excitation of coherent magnons. 
In contrast, the resonance field for the SAF is sufficiently large to allow efficient coupling of SAW and the acoustic magnons.
From the change in the linewidth of the SAW transmission peak, we estimate the coupling constant $g$ of the acoustic magnons and the SAW phonons to be $\sim$15.6 MHz.
The corresponding magnomechanical cooperativity parameter $C$ is $\sim$0.66.
These results thus show that SAF can be used as an effective media for magnon-phonon coupling and provide a playground for on-chip cavity magnomechanics.
\\

\section{Acknowledgements}
We are grateful to R. Hisatomi, T. Taniguchi, S. Nakatsuji, H. Yamaguchi and H. Okamoto for fruitful discussion. This work was partly supported by JSPS KAKENHI (Grant Number 20J20952, 20J21915, 23KJ1419, 23H05463) from JSPS, JSR Fellowship from the University of Tokyo, MEXT Initiative to Establish Next-generation Novel Integrated Circuits Centers (X-NICS).

\bibliography{reference_100923}

\end{document}